\documentclass[twoside]{dis08}
\usepackage[latin1]{inputenc}
\usepackage[dvips]{graphicx,epsfig,color}
\usepackage{wrapfig,rotating}
\usepackage{amssymb,amsmath,array}

\begin{document}
\title{An approach to fast fits of the unintegrated gluon
density}

\author{A. Knutsson$^2$, A. Bacchetta$^1$, H. Jung$^2$,  K. Kutak$^2$
%
%
\vspace{.3cm}\\
1- Jefferson Laboratory, Theory Center, 12000 Jefferson Av. Newport News, VA, 23606, USA
%
\vspace{.1cm}\\
2- DESY, Notkestrasse 85, 22603 Hamburg, Germany
}

\maketitle

\begin{abstract}
An approach to fast fits of the unintegrated gluon density has been
developed and used to determine the unintegrated gluon density by
fits to deep inelastic scatting di-jet data from HERA. The fitting
method is based on the determination of the parameter dependence by help of
interpolating between grid points in the parameter-observable space
before the actual fit is performed.
\end{abstract}

\section{Introduction}
The substructure of the proton is parameterized by parton density
functions (PDFs). In perturbative QCD the PDFs are given by solutions
of integral equations, for which the initial input distributions have
to be determined by fits. It turns out that, for exclusive final
states, it is not statistically efficient to tune Monte Carlo event
generators (MC) by sequential calls of the generator together with a
minimisation program. Motivated by~\cite{Abreu:1996na},
we use an alternative fitting method, which is based on
producing a grid in parameter-observable space. This allows the
parameter dependence to be determined by polynomial interpolation
before the fit is performed, which significantly reduce the time for
performing the fit.

Here we determine the parameters in the starting distribution of the
unintegrated gluon density function (uGDF) by using the CASCADE
Monte Carlo event generator~\cite{Jung:2001hx}. The fit is performed
to low $Q^2$ di-jet data from the H1 experiment~\cite{Aktas:2003ja}.

\section{The Unintegrated Gluon Density} 
The starting distribution of the unintegrated gluon density is
 parameterized as
\begin{equation}
A_0(x,k_t)=Nx^{-B}(1-x)^C\exp{(k_t-\mu)^2/\sigma^2}
\nonumber
\end{equation}
where $x$ is the longitudinal momentum fraction of the proton carried
by the gluon and $k_t$ its transverse momentum. Here we use the
fitting method to determine the normalisation, $N$, the small $x$
behaviour, $B$, and the shift, $\mu$, of the Gaussian for the
transverse momentum of the non-perturbative gluon. The parameters $C$
and $\sigma$, is kept fixed at $C=4$ and $\sigma=2$. 

The starting distribution is evolved to higher scales by gluon
emissions according to the CCFM evolution equation which impose
angular ordering of the emitted gluons.

\section{The Fitting Method}
In the first step of the fitting procedure we build up a grid of MC
predictions in the parameter space $(p_1,p_2,\dots,p_n)$ for each of the
observables $X$.

Secondly, we use the grid to describe the parameter-observable space
analytically, by using a
polynomial of the form
\begin{equation} 
X(p_1,p_2,..,p_n)=A_0+\sum_{i=1}^n B_ip_i+\sum_{i=1}^n C_i
p_i^2+\sum_{i=1}^{n-1}\sum_{j=i+i}^nD_{ij}p_ip_j+\textrm{H.O.} 
\nonumber
\end{equation}   
It is here possible to minimise $\chi^2$ with singular value decomposition
(SVD)\cite{NR}, since the coefficients which are to
be determined ($A_0$, $B_1$,$\dots$) are mapping the MC grid on
the observables in an over determined system of linear equations.
This is done separately for each of the MC predicted data points.

In order to account for correlations between parameters the form of
the polynomial has to be of at least second order. In
the presented fit we use a third order polynomial, which gives a
significantly better description of the parameter space. 

Having determined the polynomial describing the parameter
space, we can fit the
parameters $p_1,p_2,\dots$. This is done by applying a $\chi^2$
minimisation to
\begin{equation}
\chi^2=\sum
\frac{(X_{k,poly}-X_{k,data})^2}{(\delta X_{k,poly}-\delta X_{k,data})^2}
\nonumber
\end{equation}
where the sum runs over \textit{all data points, $k$}.
$X_{k,data}$ is the measured data point, with the corresponding experimental error
$\delta X_{k,data}$, and $X_{k,poly}$ the
polynomial prediction, with the error $\delta
X_{k,poly}$ calculated from the individual errors of the
fitted coefficients by using the covariance matrix. To perform this last
step we use MINUIT since the dependence on parameters $p_1, p_2,...$
is non-linear. 

The method turns out to be very time efficient in particular since the MC
grid points are generated simultaneously and instead of fitting MC to
data we perform a fit of the polynomials, which is much faster due to
the fact that the event generation already is performed. The method
also allows for very fast refitting if one, for example, wants to
study the exclusion of some experimental data points.
\begin{figure}
\centerline{\includegraphics[width=0.9\columnwidth]{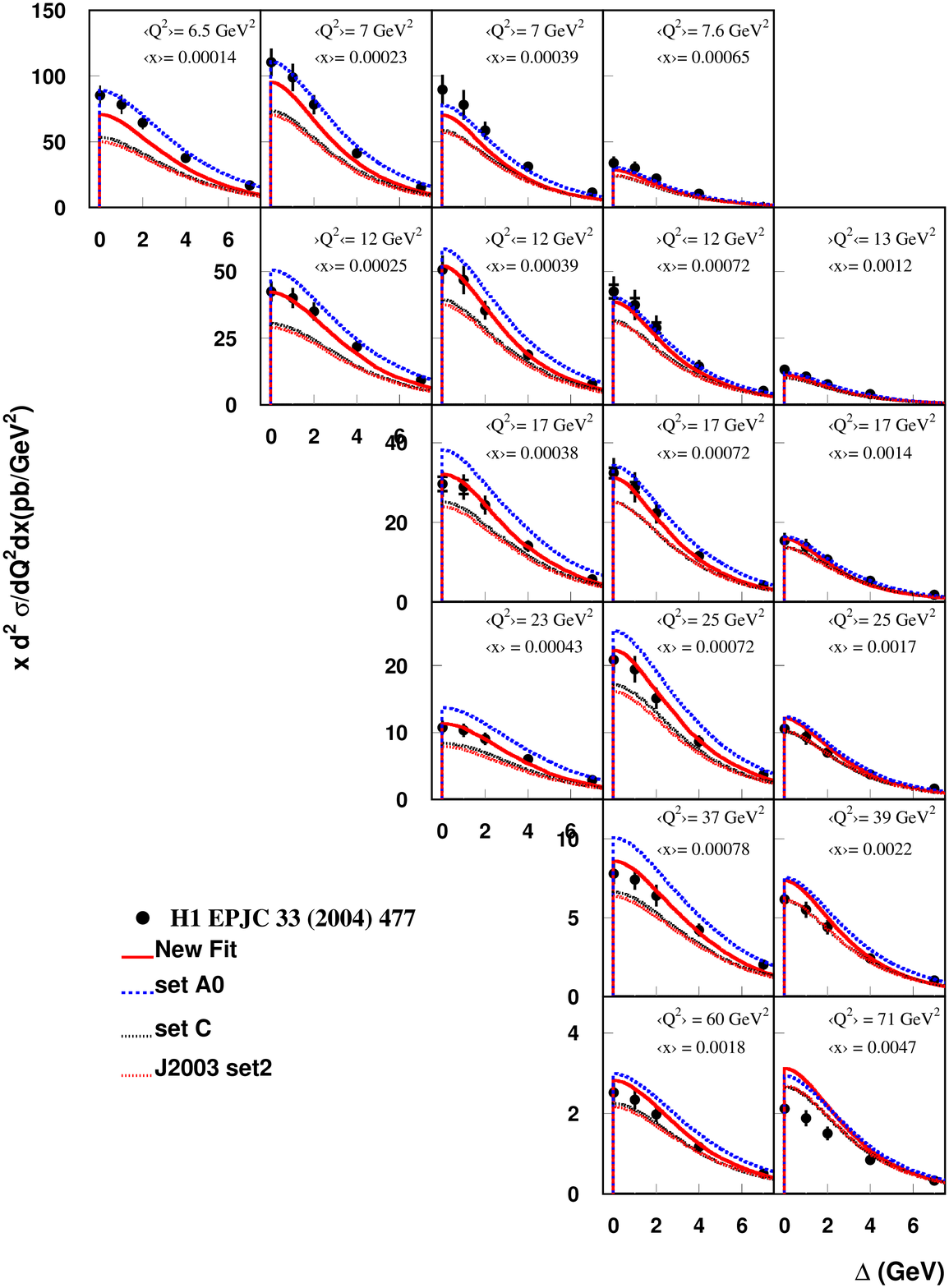}}
\caption{\label{Fig: 1}Di-jet data as a function of the difference
between the transverse momentum requirement on the di-jets, $\Delta$,
in bins of $x_{\textrm{Bj}}$ and $Q^2$ compared to predictions from
the CASCADE Monte Carlo event generator using the newly fitted PDF
(full line) and 3 old PDFs; set A0 (dashed), set C(dotted/dark),
J2003 set 2(dotted/light) .}
\end{figure}

\section{The Experimental Data} 
The measurement used for the fit was made by H1~\cite{Aktas:2003ja}
at $\sqrt{s}=318$~GeV within the kinematic range $5 < Q^2 <
100$~GeV$^2$, $10^{-4}<x_{Bj}<10^{-2}$ and $0.1<y<0.7$. Jets are
defined in the $\gamma$p rest frame with the inclusive
$k_t$-algorithm, and required to fulfill $E^{HCM}_T > 5$~GeV and $-1
< \eta^{LAB} < 2.5$, where $E^{HCM}_T$ is the transverse momentum in
the hadronic center of mass frame and $\eta^{LAB}$ is the
pseudorapidity in the laboratory frame. Events with at least two jets
fulfilling these requirements were analysed. The total di-jet
cross-section was measured as a function of $\Delta$ which gives an
additional restriction on the hardest di-jet according to $E^{HCM}_T
> 5 + \Delta$~GeV. The cross-section were made double differential
by  binning in $x_{Bj}$ and $Q^2$.

\section{Results}
\begin{figure}[h!]
\centerline{\includegraphics[width=0.3\columnwidth]{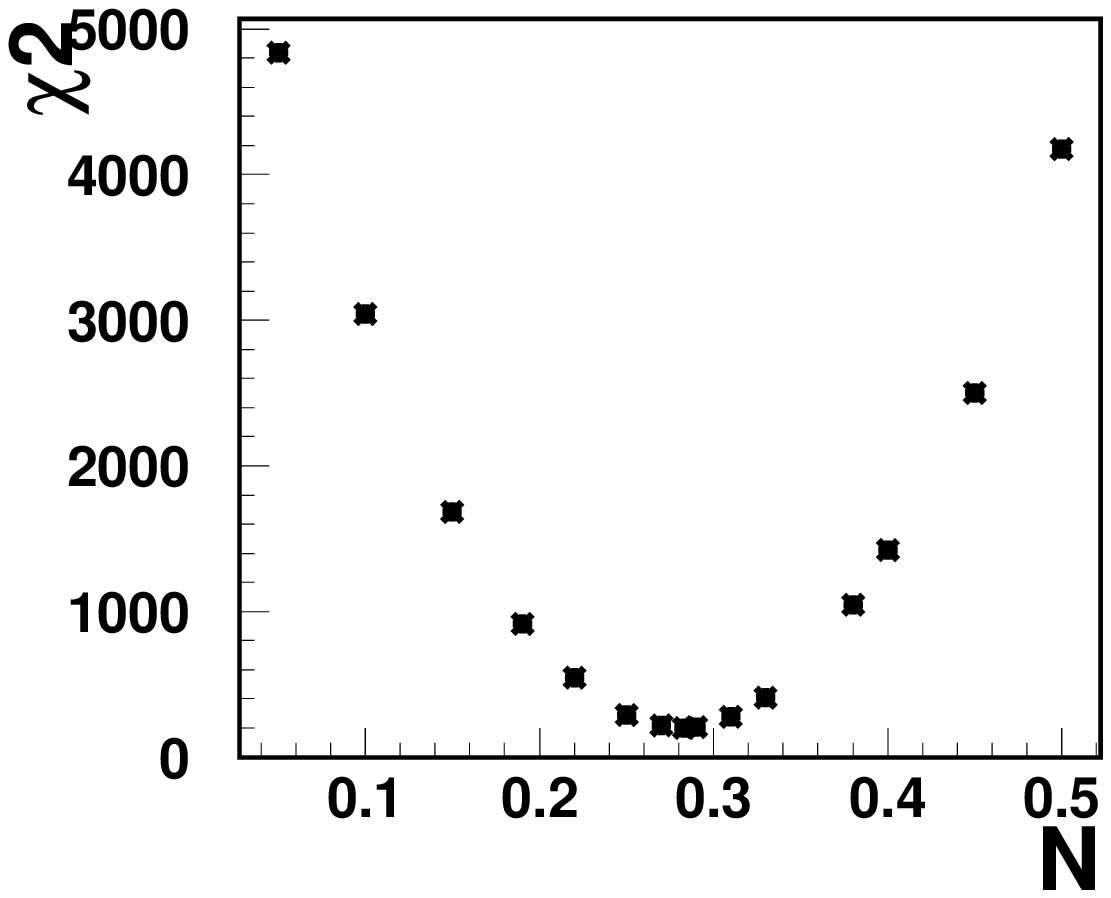}
	    \includegraphics[width=0.3\columnwidth]{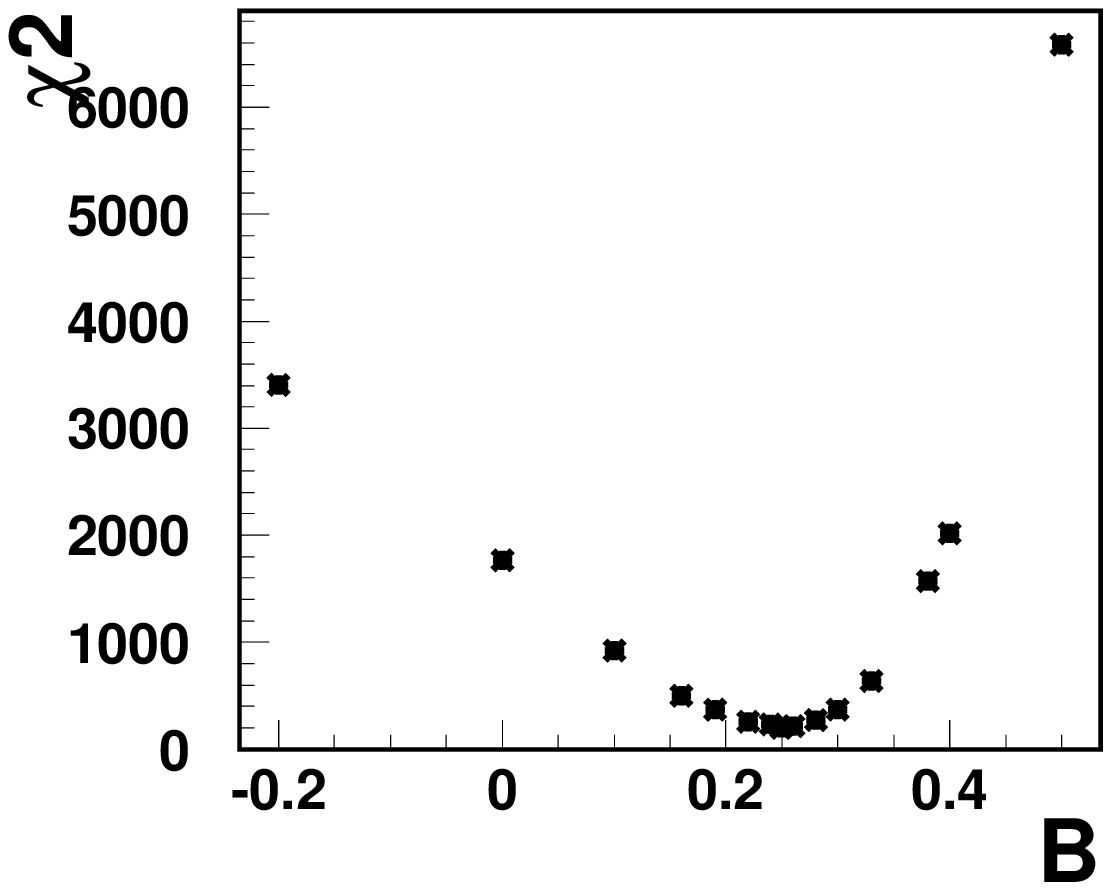}
	    \includegraphics[width=0.3\columnwidth]{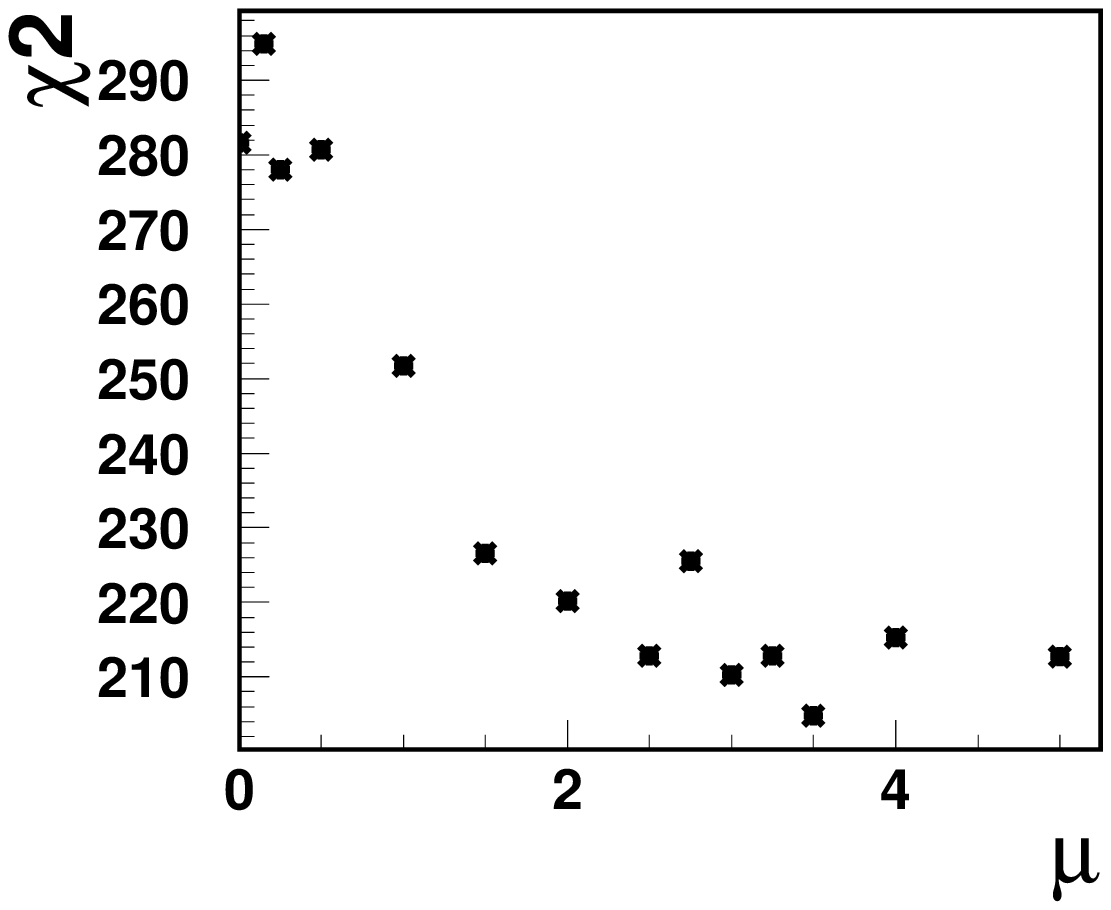}
	    }
\caption{\label{Fig: 2} $\chi^2$ profiles as a function of the fitted parameters.}
\end{figure}

The result of the fit to the H1 di-jet data is shown in
Fig.~\ref{Fig: 1} together with the data as well as three already
existing uGDFs. The parameter values determined from the fit are
$N=0.28$, $B=0.25$ and $\mu=3.0$.  Scanning $\chi^2$ as a function of
these parameters, as shown in Fig.~\ref{Fig: 2}, confirms that this
is a minimum for $N$ and $B$, and $\mu$ has flattened out at 3.
Varying this parameter around 3 thus not change the MC
predictions.

\begin{figure}[h!]
\centerline{\includegraphics[width=0.4\columnwidth]{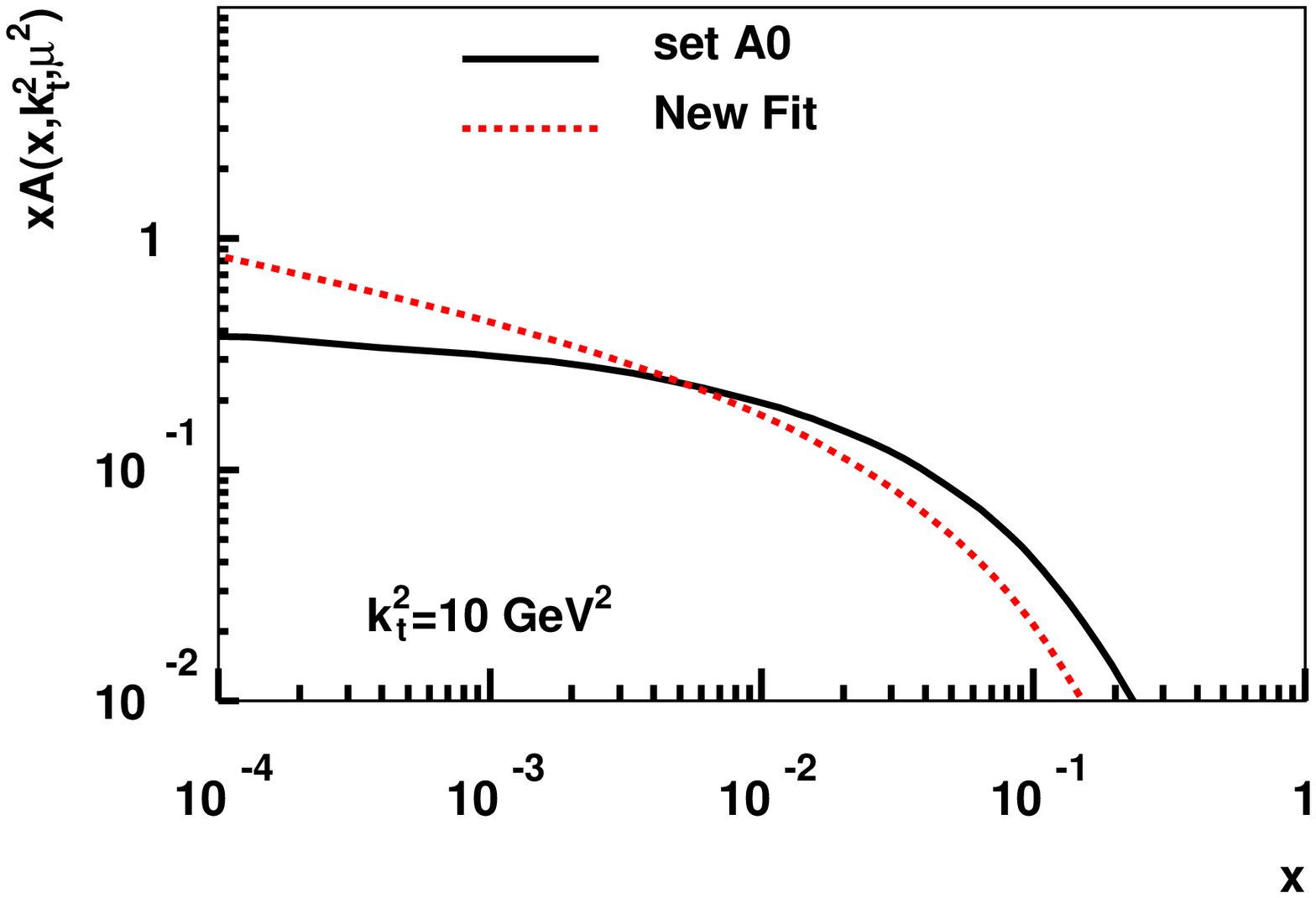}
	    \includegraphics[width=0.4\columnwidth]{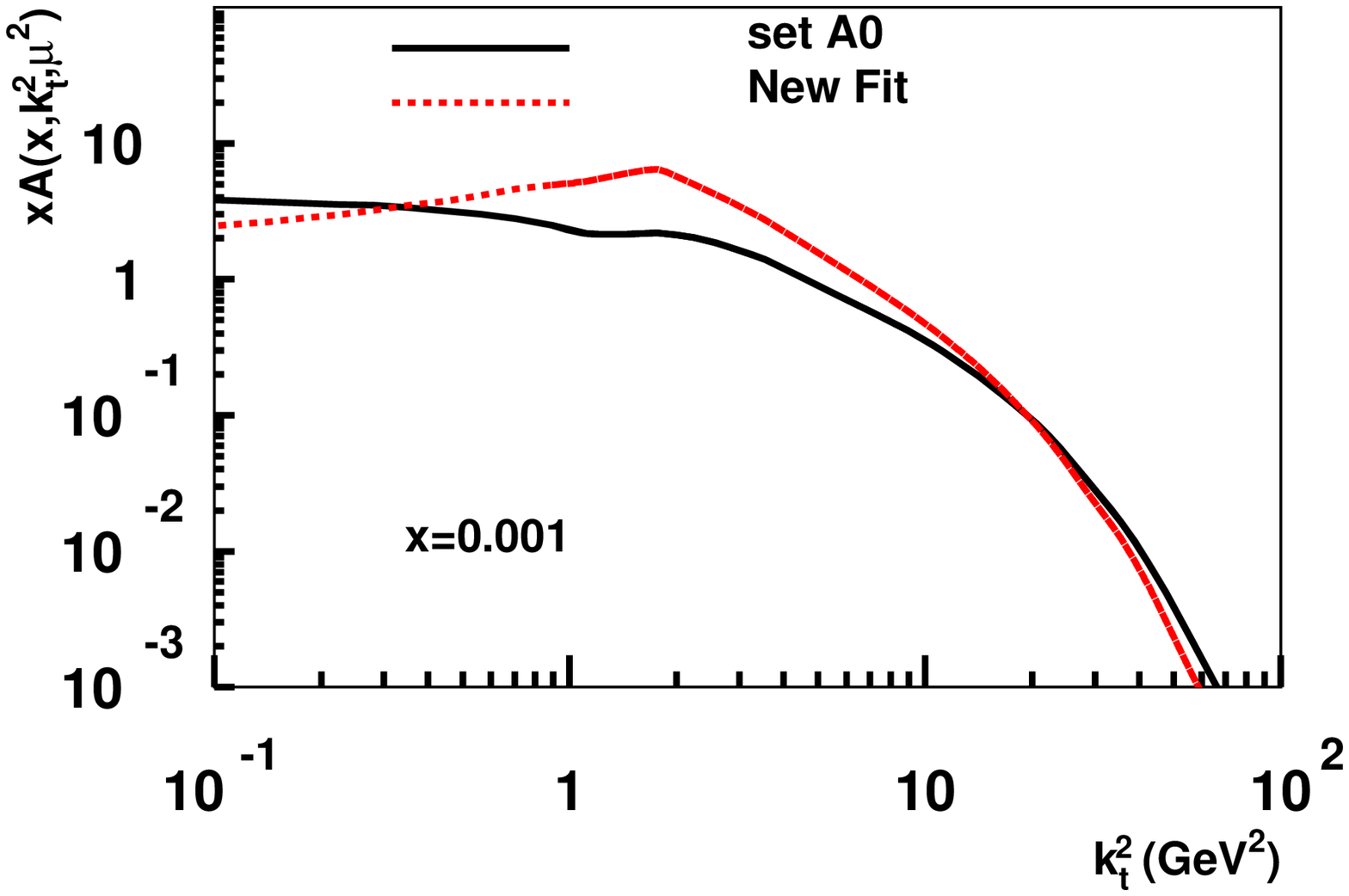}}
\caption{\label{Fig: 3} The newly fitted PDF (dashed line) compared
to the old PDF set A0 (full line), drawn as a function of $x$ (for
$k^2_t=$10 GeV$^2$) and $k^2_t$ (for x=0.001).}
\end{figure}

The $\chi^2/ndf$  of the new fit is 2.1, which should be compared to
$\chi^2/ndf=$3.5 for the best performing old uGDF, which in this case
is set A0. A uGDF that was determined from fits to the proton
structure function~\cite{Jung:2001hx}. There are two significant
differences between these two uGDFs.  The first one is the small x
behaviour given by $B$, which is 0 for set A0, but suggested to be
0.25 by the fit to di-jet data. The second difference is that, while
set A0 uses a non shifted Gaussian for the gluon $k_t$, i.e.
$\mu=0$, the di-jet data suggests a large shift resulting in a
decreasing starting distribution towards low $k_t$. The differences
are illustrated in Fig.~\ref{Fig: 3} where the uGDF is draw as a
function of $x$ and $k^2_t$.


\begin{footnotesize}


%
%

\end{footnotesize}


\end{document}